\documentstyle[prl,aps,preprint,epsf]{revtex}

\begin{document}

\title{ Fermi Liquid without Quasiparticles and Electron Spectral
        Functions of Two-Dimensional High-$T_c$ Superconductors }
\author{ V. A. Khodel and M. V. Zverev }
\address{ Russian Research Center Kurchatov Institute,
Moscow 123182, Russia }
\date{\today , Submitted to Physical Review Letters }\maketitle

\begin{abstract}
Properties of strongly correlated  two-dimensional (2D) electron
systems in solids are studied on the assumption that
these systems undergo a phase transition, called
fermion condensation, whose characteristic feature  is flattening
of the electron spectrum $\epsilon({\bf p})$. Unlike the previous models
in the present study, the decay of single-particle states is properly
taken into account. Remarkably, the value  of the topological
charge ($N=1/2$) remains unchanged, supporting the view that systems with
a fermion condensate form a separate class of Fermi liquids.
Results of calculations are found to be in qualitative agreement
with ARPES data.
\end{abstract}

\vskip 1cm

The single-particle  dynamics of Fermi systems at  near zero
temperatures $T$ is known to  depend crucially on the  index
$\nu$, that characterizes the imaginary part of the mass operator,
${\rm Im\,}\Sigma(\varepsilon\to 0)\sim\varepsilon^{\nu}$, the
energy $\varepsilon$ being measured from the chemical potential $\mu$.
In  ordinary homogeneous Fermi liquids
such as nuclear matter and liquid $^3$He, where the exclusion principle
``leads the dance",  the index $\nu$ equals 2, and the Fermi liquid can be
treated as a gas of
interacting ``immortal"  quasiparticles,
the cornerstone of standard Fermi liquid theory
(SFLT) \cite{lan}. After
many successful  years,  SFLT is currently
encountering serious difficulties in treating
normal states of 2D high-$T_c$ superconductors. The analysis of ARPES
data shows that even  around the diagonals of the Brillouin zone,
the  index $\nu$ is unity \cite{kam,john},
while in the immediate vicinity of the van Hove points (vHP),
the sharp ARPES peaks  disappear altogether
 \cite{shsch,camp,john1}.

We propose that solution of this challenging problem
is associated with fermion condensation
\cite{ks,vol,noz,physrep,sol,shag,zb,normt,kz,zkb,zkc},
a novel phase transition that generates a group of dispersionless states,
called the fermion  condensate (FC), whose energies coincide with $\mu$.
States with the FC have been uncovered \cite{ks} as  unconventional
solutions of equations of Landau theory at $T=0$. To gain insight into
the problem, consider the Landau
formula for the variation  of the  ground state energy $E_0$
\begin{equation}
\delta E_0=\sum\limits_{\bf p} \epsilon({\bf p}; n_F)\,
\delta n({\bf p})
+ {1\over 2} \sum\limits_{{\bf p}_1,{\bf p}_2}
\Gamma({\bf p}_1,{\bf p}_2; n_F)
\,\delta n({\bf p}_1)\,\delta n({\bf p}_2) \ ,
\label{lan}
\end{equation}
 applicable if the change  of the quasiparticle momentum
distribution $n_F({\bf p})=\theta(\mu-\epsilon({\bf p}))$ is rather small.
The single-particle (SP) spectrum $\epsilon({\bf p})$ and the scattering
amplitude $\Gamma({\bf p}_1,{\bf p}_2)$
are to be evaluated for the Landau state. The formula (1) is appropriate
for homogeneous systems,  where ${\bf p}$ is the momentum, as well as
for finite ones, where ${\bf p}$ stands for a set
of relevant  SP quantum numbers. The crucial point is that
due to the discreteness of the SP spectra, low-lying states of finite
systems, adjacent to the ground state, do not practically damp. This
situation  persists  until the excitation energy $\varepsilon$ attains
values $\sim \varepsilon^0_F/\sqrt{A}$, ($\varepsilon^0_F$ is the Fermi
energy, $A$ is the particle number), above which the SP widths
$\gamma_f(\varepsilon)$, calculated with incorporation of broadening
of energy $\delta$-functions, exceed the distance between
the SP levels. Then the widths $\gamma_f(\varepsilon)$ and
$\gamma(\varepsilon)$, inherent in
infinite systems, rapidly approaches each other.  We infer that in
search for a new ground state of a large system,   the momentum
representation can be employed, while
damping effects can be ignored. At the same time, in dealing with the
spectral functions  at realistic energies,   damping
effects  should be properly taken into account.

As seen from Eq.(\ref{lan}), the Landau state holds until the
necessary stability condition (NSC), requiring the positivity of the
l.h.s.
of Eq.~(\ref{lan}) at any admissible variation of
$n_F$,
is violated, and thereby the minimum of $E_0(n)$ moves  away from
$n_F$.
 A prerequisite for this to occur is  the presence
of substantial momentum-dependent components in $\Gamma$ (see below).
For a long time, this problem was studied only  under the assumption
that the new Fermi
surface becomes multi-connected, while the  occupation numbers
$n({\bf p})$ remain 1 or 0 \cite{llano}. However, the expansion (\ref{lan})
can also be applied to evaluate  variations of $E_0$ when continuous
{\it quasiparticle} distributions $ n({\bf p})$ come into play,  and
beyond the point where the NCS is violated, it
often turns out that such unusual for SFLT, but accustomed e.g. for
the Bogoliubov theory of superconductivity, distributions  yield
a lower energy than any discrete-valued
solution with  multi-connected Fermi surface. This compels us to treat
the quasiparticle occupation numbers $n({\bf p})$ in Eq.~(\ref{lan})
as variational
parameters  and look for a new location of the minimum of $E(n)$
 from the variational condition
\cite{ks}
\begin{equation}
{\delta E_0(n)\over\delta n( {\bf p})}=\mu\  , \qquad
{\bf p}\in {\rm C} \  ,
\label{var}
\end{equation}
with $\delta E_0$ given by Eq.~(\ref{lan}). The region C is the place
 of the residence of the fermion condensate, since
 the l.h.s. of Eq.~(\ref{var}) is just
 the quasiparticle energy $\epsilon({\bf p};n)$, and
Eq.~(\ref{var}) is equivalent to $\epsilon({\bf p}\in {\rm C};n)=\mu $.
Outside this domain, whose boundaries are determined from
Eq.~(\ref{var}) itself, the old solution $n_F({\bf p})$ survives.

Similar results can be obtained by summation of  diagrams of the mass
operator in Dyson equation \cite{zkc}. The transition occurs, if
in this equation, there exists a bifurcation point. Beyond this point,
the ground state distribution $n({\bf p})$ is evaluated from relation
\begin{equation}
\epsilon^0_{{\bf p}}+\Sigma({\bf p},\varepsilon=0;n)= \mu \  ,
\qquad {\bf p}\in {\rm C}  \  ,
\label{an1}
\end{equation}
 which is equivalent to the equation $\epsilon({\bf p}\in {\rm C};n)=\mu$.
If one neglects for a while  the $\varepsilon$-dependence of
$\Sigma({\bf p},\varepsilon)$, then  the FC Green function
$G({\bf p}\in {\rm C},\varepsilon)=[\varepsilon+\mu-\epsilon^0_{{\bf p}}-
\Sigma({\bf p}\in {\rm C},\varepsilon)]^{-1}$ becomes
 $1/\varepsilon$,
that leads to    a change of the Fermi
liquid topological charge $N$ \cite{vol}.
Recall that for ordinary homogeneous Fermi liquids, with
$G( p,\varepsilon)\sim [\varepsilon-\epsilon(p)+\mu]^{-1}$,
and for
Luttinger liquids as well, the value of $N$ is unity, as given
by the integral
\begin{equation}
N=\oint\limits_L {dl\over 2\pi i}\,G({\bf p},\varepsilon{=}i\Omega)\,
\partial_l G^{-1}({\bf p},\varepsilon{=}i\Omega) \  ,
\label{top}
\end{equation}
taken along the contour $L$, embracing the Fermi line in the
3D space $(p_x,p_y,\Omega)$. However, upon inserting
$G_{\mbox{\scriptsize C}}( p,\varepsilon)= 1/\varepsilon$
into the integral (\ref{top}), one finds  $N=1/2$, implying that
systems with the FC form a separate class of normal Fermi liquids
\cite{vol}. As we shall see, this result holds if the energy
dependence of the mass operator is incorporated.

The substantial momentum dependence of $\Gamma$, needed
for solution of Eqs.~(\ref{var}) and ~(\ref{an1}) to exist,
emerges, for example, in the vicinity of an impending
antiferromagnetic phase
transition. In this case, the static spin susceptibility
    $\chi({\bf q},\omega=0)
\sim [\beta^2+\kappa^2({\bf q}-{\bf Q})^2]^{-1}$ \cite{pines}
 diverges at ${\bf q}\to {\bf Q}=(\pi,\pi)$,
since $\beta$ vanishes at the transition point.
The direct  term $\Gamma_d$ in $\Gamma$, being proportional to
$-\chi({\bf q},\omega=0)$, diverges as well, and so does the exchange
term $\Gamma_e({\bf p},{\bf p}_1,{\bf q})\sim
[\beta^2+\kappa^2({\bf p}_1-{\bf p}_2
+{\bf q}-{\bf Q})^2]^{-1}$ because of the
antisymmetry relations imposed on $\Gamma$ \cite{dug}. It is the
exchange term,
taken at ${\bf q}=0$,  that enters Eq.~(\ref{lan}). Parameters
$\beta$ and $\kappa$ are assumed to depend on  the density $\rho$
rather than on $n_F({\bf p})$ itself,  so that
$\delta\Gamma_e/\delta n_F({\bf p})=0$. If the amplitude $\Gamma$
is approximated by $\Gamma_e$, then  upon inserting Eq.~(\ref{lan})
into Eq.~(\ref{var}) the latter is recast to
$\epsilon_1({\bf p};n_F) + \sum\limits_{{\bf p}_1}
\Gamma_e({\bf p},{\bf p}_1;n_F) \, n({\bf p}_1)= \mu$.
From the definition $\epsilon_1({\bf p})= \epsilon({\bf p};n_F) -
\sum\limits_{{\bf p}_1}\Gamma_e({\bf p},{\bf p}_1)n_F({\bf p}_1)$,
it is clear that the variational
derivative  $\delta \epsilon_1({\bf p})/\delta n_F({\bf p})$
vanishes. Therefore $\epsilon_1({\bf p})$
is reduced to the LDA spectrum $\epsilon^0_{{\bf p}}$, and
Eq.~(\ref{var})  takes the  form
\begin{equation}
\epsilon^0_{{\bf p}} + \sum\limits_{{\bf p}_1}
\Gamma_e({\bf p},{\bf p}_1) \, n({\bf p}_1) = \mu \ , \qquad {\bf p}\in
{\rm C}
\label{sps}
\end{equation}
often employed in theory of fermion condensation. It is worth noting
that Eq.~(\ref{sps})
can be derived from the effective  energy functional
$E_{\rm eff}(n)=\sum\limits_{{\bf p}}\epsilon^0_{{\bf p}}\,
n({\bf p})+1/2 \sum\limits_{{\bf p}{\bf p}_1}
\Gamma_e({\bf p},{\bf p}_1)\,n({\bf p})\,n({\bf p}_1)$, which is
appropriate only for the rearrangement problem and
 has no connection with the Hartree-Fock part of
 the  ground-state energy functional.

Results of numerical calculations of Eq.~(\ref{sps}), which are
insensitive to the choice of the parameter $\kappa$, show that
a FC arises when the filling approaches $1/2$.
Once it appears, the FC resides close to the vHP. As seen from Fig.~1,
the fraction $\eta$ of the Brillouin zone occupied
by the FC remains rather small, attaining a maximum
$\sim 0.1$ when the filling slightly exceeds 1/2,  and the hole
pocket is centered around  $(\pi,\pi)$ (for more details, see \cite{zkc}).

The degeneracy of the SP spectrum at $T=0$, a salient feature
of the solution given by Eqs.~(\ref{var}), (\ref{an1}), is lifted
by pairing interactions which are not included into
Eq.~(\ref{lan}) \cite{ks,zkc}. In doing so,  the BCS occupation
numbers $v^2({\bf p})$
coincide with  $n({\bf p})$ evaluated from Eq.~(\ref{var}) or
Eq.~(\ref{an1}) provided the BCS coupling constant is
rather small. Evidently, in obtaining both these
distributions damping effects can be ignored. For this reason,
superfluid systems with  and without the FC look more alike than
normal ones, since in normal states of conventional Fermi liquids
the damping  makes no difference, whereas in normal states of
systems with the FC,  the damping becomes  a real ``weathermaker".
Indeed,  the relevant contribution to
${\rm Im}\,\Sigma_R({\bf p},\varepsilon)$ is given by \cite {trio}
$$
{\rm Im}\,\Sigma_R({\bf p},\varepsilon)\sim
\sum\limits_{{\bf q},{\bf p}_1}
\int\!\!\int d\omega\, d\varepsilon_1 F(\varepsilon,\omega,\varepsilon_1,T)\,
|\Gamma({\bf p},\varepsilon,{\bf p}_1,\varepsilon_1,{\bf q},\omega;n)|^2
$$
\begin{equation}
\times\, {\rm Im}\,G_R({\bf p}-{\bf q},\varepsilon-\omega)\,
{\rm Im}\,G_R(-{\bf p}_1,-\varepsilon_1)\,
{\rm Im}\,G_R({\bf q}-{\bf p}_1,\omega-\varepsilon_1)\  ,
\label{immas}
\end{equation}
where the factor
$F(\varepsilon,\omega,\varepsilon_1,T)=\cosh({\varepsilon\over 2T})
[\cosh({\varepsilon_1\over 2T})\cosh({\varepsilon-\omega\over 2T})
\cosh({\omega-\varepsilon_1\over 2T})]^{-1}$,
$|\Gamma|^2$ is the absolute square of the scattering amplitude,
properly averaged over spin variables, and $G_R$ is the retarded
Green function with
${\rm Im}\,G_R({\bf p},\varepsilon)=
-\gamma({\bf p},\varepsilon) [(\varepsilon-\sigma({\bf p},\varepsilon)-
s({\bf  p}))^2+\gamma^2({\bf p},\varepsilon)]^{-1}$, where
$\sigma({\bf p},\varepsilon)
={\rm Re}\,\Sigma_R({\bf p},\varepsilon)-
{\rm Re}\,\Sigma_R({\bf p},\varepsilon=0)$
and $ s({\bf p})=
\epsilon^0_{{\bf p}}+{\rm Re\,}\Sigma_R({\bf p},\varepsilon=0)-\mu$.

We restrict ourselves to temperatures $T$ markedly lower than
the characteristic temperature $T_f$
of destroying the FC that allows one to ignore the $T$-dependence
of $\eta$. To simplify the problem, we replace the function
$\gamma({\bf p},\varepsilon)$ by a set of functions of the single
variable $\varepsilon$, e.g.  in the FC region,
$ \gamma({\bf p}\in {\rm C},\varepsilon)$ is reduced to
$\gamma_{\mbox{\scriptsize C}}(\varepsilon)$.
The complementary
region of momentum space, in which the dispersion of the spectrum
$\epsilon({\bf p})$ has a nonzero value, is composed of
two subregions.  The first, adjacent to the FC domain and henceforth
denoted by T, is a transition region, in which the same decay processes,
as in the FC region, are still kinematically allowed.  The second
subregion,  denoted by M, is located around diagonals of the Brillouin
zone. Here some of these processes are either kinematically forbidden
or at least strongly suppressed. Correspondingly,
$\gamma({\bf p}\in {\rm T},\varepsilon)\to
\gamma_{\mbox{\scriptsize T}}(\varepsilon),
\gamma({\bf p}\in {\rm M},\varepsilon)\to
\gamma_{\mbox{\scriptsize M}}(\varepsilon)$. To close the set of
equations of the problem, the
amplitude $\Gamma$ should somehow be specified. Bearing in mind that
$\eta$ is small, we shall initially neglect the FC contributions
to $\Gamma$, replacing it by $\Gamma(n_F)$.

We start with the case $\varepsilon\gg T$, and  set $T=0$ in the
integral (\ref{immas}), thus dropping all  $T$-dependent contributions.
First we evaluate
$\gamma_{\mbox{\scriptsize C}}(\varepsilon\to 0)$. In this case,
(i) contributions   from
 processes involving only  the FC states prevail (see below), (ii)
the quantity $|\Gamma(n_F)|^2\sim \beta^{-4}$ can be factored out of
the integral (\ref{immas}), and (iii)  the  quantity
$ s({\bf p}\in {\rm C},T)$, which
vanishes  over the FC region at $T=0$, can be verified to remain
small compared to leading terms in Eq.~(\ref{immas}),
and thus can be  neglected. As a result, the energy and momentum
integrations in Eq.~(\ref{immas})  separate. Taking for certainty
$\varepsilon> 0$ and  omitting numerical factors, we are left with
\begin{equation}
\gamma_{\mbox{\scriptsize C}}(\varepsilon\to 0)\sim
\beta^{-4}\eta^2\int\limits_0^{\varepsilon} \! \int\limits_0^{\omega}
A_{\mbox{\scriptsize C}}(\varepsilon-\omega)\,
A_{\mbox{\scriptsize C}}(-\varepsilon_1)\,
A_{\mbox{\scriptsize C}}(\omega-\varepsilon_1)\,
d\omega\, d\varepsilon_1 \  ,
\label{dc}
\end{equation}
where $A_{\mbox{\scriptsize C}}(\varepsilon)=
{\rm Im}\,G_R({\bf p}\in {\rm C},\varepsilon)$. To proceed,
we insert
$\gamma_{\mbox{\scriptsize C}}(\varepsilon\to 0)
\sim \varepsilon^{\nu_{\mbox{\tiny C}}}$
into the Kramers-Kr\"onig relation to obtain
$\sigma_{\mbox{\scriptsize C}}(\varepsilon\to 0)
\sim\varepsilon^{\nu_{\mbox{\tiny C}}}$. We then substitute
$\gamma_{\mbox{\scriptsize C}}$ and $\sigma_{\mbox{\scriptsize C}}$
into  $A_{\mbox{\scriptsize C}}$ and find
$A_{\mbox{\scriptsize C}}(\varepsilon\to 0)\sim
\varepsilon^{-\nu_{\mbox{\tiny C}}}$.
Finally, upon inserting this result into Eq.~(\ref{dc}), we arrive at
$\nu_{\mbox{\scriptsize C}}=1/2$ \cite{kz}. More precisely, one obtains
$\gamma_{\mbox{\scriptsize C}}(\varepsilon\to 0)
\sim \beta^{-1}(\eta\varepsilon^0_F\varepsilon)^{1/2}$  and
\begin{equation}
G_R({\bf p}\in {\rm C},\varepsilon\to 0)\sim e^{-i\pi/4}
[\gamma_{\mbox{\scriptsize C}}(\varepsilon\to 0)]^{-1}\sim
      e^{-i\pi/4}\beta(\eta\varepsilon^0_F\varepsilon)^{-1/2} \ .
\label{gr}
\end{equation}
This result can be shown
to hold even if the momentum dependence of the quantities
$\gamma_{\mbox{\scriptsize C}}({\bf p},\varepsilon)$ and
$\sigma_{\mbox{\scriptsize C}}({\bf p},\varepsilon)$
is properly taken into account.
We see that in the FC region,
 the conventional structure of the Green
function  is destroyed, the  familiar pole being
 replaced by a branch point at $\varepsilon=0$.
What happens to the topological charge $N$?
 Upon inserting the Green function (\ref{gr}) into Eq.~(\ref{top})
 and performing
simple integration, we are again led to  Volovik's previous  result
$N=1/2$ \cite {vol},  in spite of the dramatic alteration of the
Green function itself that occurs in the FC region.

In the transition region T,  the decay  into the FC states
is not kinematically forbidden. Accordingly,
$\gamma_{\mbox{\scriptsize T}}(\varepsilon\to 0)\sim
\beta^{-1}(\eta\varepsilon^0_F\varepsilon)^{1/2}$,
while the function
$s({\bf p}\in {\rm T})$
already differs from zero. Requiring it to vanish
at the   boundaries of the FC region along with its
first derivative, one finds that in the region T, the
conventional structure of the Green function  is recovered,
but  in the vicinity of the FC domain, single-particle excitations
appear to be ill-defined, since the pole of $G({\bf p},\varepsilon)$
is located  close to the imaginary energy axis.

In the region M,    dominant
contributions to ${\rm Im}\,\Sigma_R(\varepsilon)$ come from a process
associated with  the generation of three states: two from the
FC region and  one from the M region. In this case, the formula
for finding  $\gamma_{\mbox{\scriptsize M}}(\varepsilon\to 0)$ reads
\begin{equation}
\gamma_{\mbox{\scriptsize M}}(\varepsilon{\to}0)\sim \beta^{-4}
\sum\limits_{{\bf p},{\bf p}_1}
\int\limits_0^{\varepsilon}\!\int\limits_0^{\omega}
A_{\mbox{\scriptsize C}}(-\varepsilon_1)\,
A_{\mbox{\scriptsize C}}(\omega{-}\varepsilon_1)\,
[1-\theta({\bf p})]\,
P_{\mbox{\scriptsize C}}({\bf p}{-}{\bf p}_1)\,
A_{\mbox{\scriptsize M}}({\bf p},\varepsilon{-}\omega)\,
d\omega\, d\varepsilon_1 \  ,
\label{dn}
\end{equation}
where $P_{\mbox{\scriptsize C}}({\bf q})=\sum_{{\bf p}}\theta({\bf p})
\theta({\bf p}-{\bf q})$ and
 $\theta({\bf p})=1$  if ${\bf p}\in {\rm C}$ and
otherwise vanishes. It is seen that in this case, the
momentum and energy integrations do not  separate.
However, one can take advantage of the fact that the spectrum
$\xi_{\mbox{\scriptsize M}}({\bf p})=
\epsilon_{\mbox{\scriptsize M}}({\bf p})-\mu$ is proportional to
$(p-p_F)$ and introduce $\xi_{\mbox{\scriptsize M}}({\bf p})$ as a new
variable.  Then after simple integration, we are
led to the result $\nu_{\mbox{\scriptsize M}}=1$ postulated in the
model of a marginal Fermi liquid (MFL) ~\cite{varma}.
Evaluation of the $\eta$-dependence of relevant quantities in
the M region yields
$\gamma_{\mbox{\scriptsize M}}(\varepsilon\to 0)\sim
\beta^{-2}\eta^{1/2}\varepsilon$
and
$\sigma_{\mbox{\scriptsize M}}(\varepsilon\to 0)\sim
\beta^{-2}\eta^{1/2}\varepsilon\ln|\varepsilon| $.

These results can be applied to the case
$\varepsilon\sim T$, where according to Eq.~(\ref{gr}),
the leading term in the FC Green function has the form
$G_R({\bf p}\in {\rm C},\varepsilon) \sim
e^{-i\pi/4}[\gamma_{\mbox{\scriptsize C}}(\varepsilon,T)]^{-1}$.
Upon inserting this expression into Eq.~(\ref{dc}) one finds that
the damping
$\gamma_{\mbox{\scriptsize C}}(x,T)$, where $x=\varepsilon/T$,
can be displayed as $\gamma_{\mbox{\scriptsize C}}(x,T)=
\gamma_{\mbox{\scriptsize C}}\sqrt{T\varepsilon^0_F}\,D(x)$ where
the constant $\gamma_{\mbox{\scriptsize C}}$ specifies
the compound, while  the  dimensionless quantity $D(x)$ obeys
the  integral  equation
\begin{equation}
D(x)= \cosh {x\over 2}
\int\!\!\int {dy\, dz\over \cosh {y\over 2}\cosh{x{-}z\over 2}
\cosh{z{-}y\over 2}\,
D(-y)\, D(x{-}z)\,D(z{-}y)} \  .
\label{dimc}
\end{equation}
With this result, the damping $\gamma_{\mbox{\scriptsize M}}(x,T)$
is  calculated  straifghtforwardly:
\begin{equation}
\gamma_{\mbox{\scriptsize M}}(x,T)=
\gamma_{\mbox{\scriptsize M}}T\cosh {x\over 2} \int\!\!\int
{dy\, dz\over  \cosh {y\over 2}\cosh{x{-}z\over 2}\cosh{z{-}y\over 2}\,\,
 D(-y)\,D(z{-}y)} \ ,
\label{dimm}
\end{equation}
the constant $\gamma_{\mbox{\scriptsize M}}$ being a characteristics
of the given compound. The function
$\gamma_{\mbox{\scriptsize M}}(x,T)/T$ starts out of the origin
as a parabolic function
    $\gamma_{\mbox{\scriptsize M}}(x,T)/T\sim 1+0.1\,x^2$.
The asymptotic
regime $\gamma_{\mbox{\scriptsize M}}(x,T)/T \sim x$, stemming from
Eq.~(\ref{dn}), begins at $x\sim 2.5$.

In normal states  with the nonzero value of the pseudogap, relations
(\ref{dimc}), (\ref{dimm}) hold, as long as it remains less than
the damping $\gamma_{\mbox{\scriptsize C}}(T)$. These relations
persist even in superfluid states provided the gap value meets
the same restriction. On the other hand, they are violated
if  energy attains values,  at which contributions to
$\gamma(\varepsilon)$ that were omitted from Eqs.~(\ref{dc})
and (\ref{dn}) become comparable to the  terms that were retained.
A leading correction
$\delta\gamma_{\mbox{\scriptsize C}}(\varepsilon)$ to the integral
(\ref{dc})
comes from final states, that involve   one  hole (particle) belonging
to the region T.  Eq.~(\ref{dn}) can be employed
to estimate this contribution,  with the single replacement
$s({\bf p}\in{\rm M})\to s({\bf p}\in{\rm T})$.
We find $\delta\gamma_{\mbox{\scriptsize C}}(\varepsilon\to 0)\sim
\beta^{-1}\varepsilon$, which is independent of the $\eta$ value.
Estimation of  other corrections to $\gamma(\varepsilon)$
is  carried out along the same lines, justifying the identification
of (\ref{dc}) and  (\ref{dn}) as  paramount contributions to
${\rm Im\,}\Sigma_R({\bf p},\varepsilon\to 0)$  until
$\varepsilon$ exceeds
the characteristic FC energy $\varepsilon_{FC}\simeq\eta\varepsilon^0_F$,
evaluated by comparison of
$\delta\gamma_{\mbox{\scriptsize C}}(\varepsilon)$ and
$\gamma_{\mbox{\scriptsize C}}(\varepsilon) \sim
\beta^{-1} (\eta\varepsilon^0_F\varepsilon)^{1/2}$.

At energies $\varepsilon\geq \varepsilon_{FC}$, the corrections exhibit
themselves  in full force, so
Eq.~(\ref{immas}) should be solved numerically in conjunction with
the Kramers-Kr\"onig relation,
employed to connect $\gamma(\varepsilon)$ and $\sigma(\varepsilon)$.
This is done with the aid of an iteration procedure,  which
converges rapidly. In Fig.~2 we display results from these
calculations  carried out with the same amplitude
$\Gamma$ that was employed
in the calculations of the FC characteristics shown in Fig.~1.
Two different $\eta$  values, specifying the fraction of
the Brillouin zone occupied by the FC,  were considered:
(a) $\eta=0.1$, close to the maximum $\eta$ value
in the model of fermion condensation driven
by antiferromagnetic fluctuations \cite{zkc}, and (b) $\eta=0.01$.
In spite of the simplicity of  the interaction adopted, salient
features of ARPES data \cite{john,john1,normt} are reproduced,
including the universal behavior of the ratio
${\rm Im\,}\Sigma_R({\bf p}\in{\rm M},\varepsilon,T;\eta)/T$
as a function of $x=\varepsilon/T$, uncovered in \cite{john}.
Moreover, our theory predicts the same behavior
of
${\rm Im\,}\Sigma_R({\bf p}\in{\rm M},\varepsilon,T;\eta)/T$ for
different compounds provided results are properly normalized.
We choose this normalization to ensure the same slope
$k(\eta)=
|\partial\,{\rm Im\,}\Sigma_R({\bf p}\in{\rm M},\varepsilon,T;\eta)/
\partial\varepsilon|$
at $x\gg 1$
where the damping changes linearly
with $\varepsilon$. The latter is demonstrated in Fig.~3, where
two functions
$|{\rm Im\,}\Sigma_R({\bf p}\in{\rm M},\varepsilon,T;\eta)|/T$,
evaluated at $\eta=0.1$ and $\eta=0.01$ and normalized to the same
slope $k=0.75$, are compared to data measured in \cite{john} for
the optimally doped cuprate Bi$_2$Sr$_2$CaCu$_2$O$_{8+\delta}$.
At the same time as follows from Eq.~(\ref{dimc}), in the FC region
the above universal scaling of the ratio
${\rm Im\,}\Sigma_R(\varepsilon,T)/T$ is destroyed, and instead
${\rm Im\,}\Sigma_R({\bf p}\in{\rm C},\varepsilon,T)$ displays
$\sqrt{T}$-dependence at $x\leq 1$.

The above scenario in which the fermion
condensation precedes the antiferromagnetic phase transition  does apply
in the three-dimensional case, although the range of the FC region
shrinks markedly. Along the same lines, one can analyze the situation
with fermion condensation in the vicinity of other second order phase
transitions, such as charge-density-wave instability.
So far the feedback of the FC on the scattering amplitude $\Gamma$
has been  ignored. However, the simplest FC diagram, i.e. a loop,
evaluated with the FC Green function (\ref{gr}), diverges
logarithmically.  As a result,  we are led to a familiar problem
of the parquet-diagram summation, solution of which
will be reported in a separate paper.

Summing up the  results of our analysis, we infer that
infinite electron systems
with a fermion condensate, independently of the dimensionality,
do not admit Landau quasiparticles, since the renormalization factor
$z=(1-\partial\Sigma/\partial\varepsilon)^{-1}_F$ that
determines the quasiparticle weight in the single-particle state
vanishes  in all the regions of the Brillouin zone. In the FC domain,
the value of the topological charge $N$ is found to be $ N=1/2$
indicating that systems  with
a fermion condensate form a separate class of Fermi liquids.
The model of  fermion condensation presented here
allows one to explain basic features of the spectral functions
of normal states of high-$T_c$ superconductors,
including the MFL behavior of the damping of SP states around
the diagonals of the Brillouin zone (the {\rm M} region) and
the suppression of the peaks in APRES data in the immediate vicinity
of the van Hove points. And the universal behavior of the ratio
${\rm Im\,}\Sigma_R({\bf p}\in{\rm M},x,T)/T$ as a function of
$x=\varepsilon/T$ established in \cite{john} for the {\rm M} region
is also reproduced in this model. Moreover,
when properly normalized, our results for
${\rm Im\,}\Sigma_R({\bf p}\in{\rm M},x,T)/T$ calculated for different
compounds collapse on the same curve.
However, as we have seen, this universal behavior is destroyed
in the {\rm C} region close to the van Hove points.

We acknowledge  P.~W.~Anderson, J.~C.~Campuzano, J.~W.~Clark,
L.~P.~Gor'kov, K.~A.~Kikoin, G.~Kotliar,
 A.~J.~Millis, J.~Mesot, N.~Nafari, M.~R.~Norman,
 E.~E.~Saperstein, and
 G.~E.~Volovik  for many stimulating discussions.
This research was supported in part by NSF Grant PHY-9900713,
by the McDonnell Center for the Space Sciences (VAK) and by the
Russian Fund for Fundamental Research,  Grant No.~00-15-96590.
VAK is grateful to  J.~W.~Clark for  kind hospitality
at Washington University in St.Louis.

\newpage

\begin{figure}
\epsfxsize=16.5cm
\epsfysize=22.cm
\centerline{\epsffile{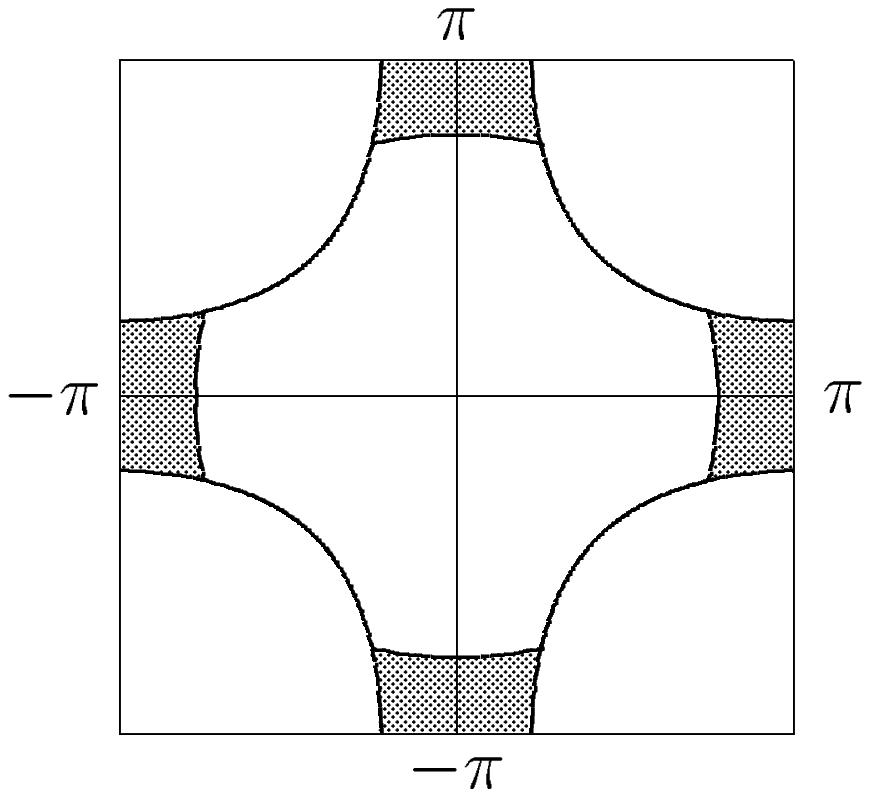}}
\vspace{-8 cm}
\caption[]{
The Fermi surface in the model of  fermion condensation,
driven by antiferromagnetic fluctuations with the scattering amplitude
$\Gamma({\bf p},{\bf p}_1)= (N(0))^{-1}
[\beta^2+\kappa^2({\bf p}-{\bf p}_1-{\bf Q})^2]^{-1}$, where $N(0)$
is the density of states and $\beta=0.2$. The regions occupied by FC
are shaded.
}
\end{figure}

\newpage

\begin{figure}
\epsfxsize=16.5cm
\epsfysize=22.cm
\hspace*{-2.5 cm}\epsffile{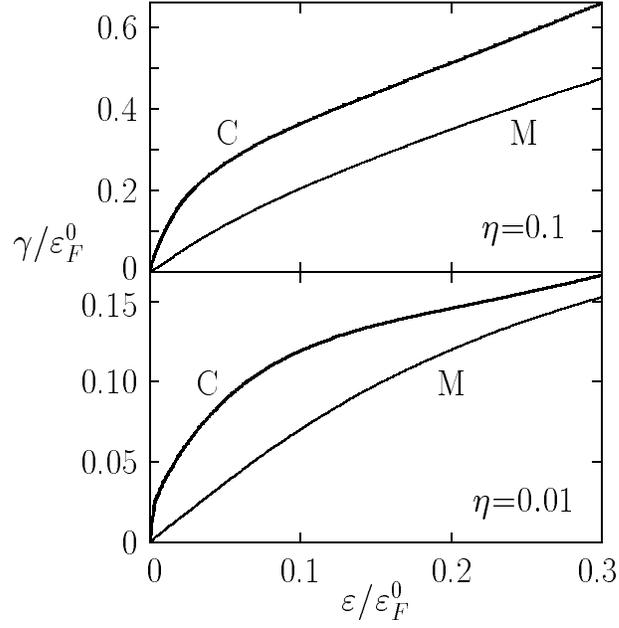}
\vspace{-7 cm}
\caption[]{
The damping of the single-particle states at $T=0$
in the vicinity of the vHP (solid lines) and around the diagonals
of the Brillouin zone (thin lines), calculated with the same scattering
amplitude $\Gamma$ for $\eta=0.1$ (upper panel) and $\eta=0.01$
(lower panel), and measured in $\varepsilon^0_F$.
}
\end{figure}

\newpage

\begin{figure}
\epsfxsize=16.5cm
\epsfysize=22.cm
\hspace*{-2.5 cm}\epsffile{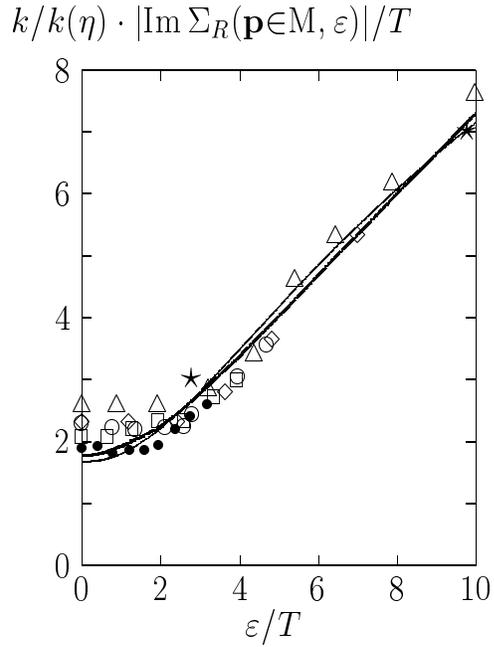}
\vspace{-7 cm}
\caption[]{
The ratio $|{\rm Im\,}\Sigma_R(\varepsilon)|/T$ around the diagonal
of the Brillouin zone as a function of
$\varepsilon/T$, calculated for $\eta=0.1$ (solid line) $\eta=0.01$
(thin line) and normalized to the same slope
$k=0.75$, i.e. multiplied by the factor $k/k(\eta)$.
The experimental data for the optimally doped cuprate
Bi$_2$Sr$_2$CaCu$_2$O$_{8+\delta}$ \cite{john} are shown by
open and solid circles ($T=300$\,K), triangles and squares
($T=90$\,K), and diamonds and stars ($T=48$\,K).
}
\end{figure}

\end{document}